# Science results from the imaging Fourier transform spectrometer SpIOMM

L. Drissen[*a], A.-P. Bernier[a], M. Charlebois[a], É. Brière[a],
C. Robert[a], G. Joncas[a], P. Martin[b], F. Grandmont[c]

[a]Dépt. de physique, de génie physique et d'optique, Université Laval, Québec, Qc, Canada G1K 7P4
and Centre de recherche en astrophysique du Québec (CRAQ)
[b]CFHT corporation, 65-1238 Mamalahoa Hwy, Kamuela, Hawaii 96743, USA
[c]ABB Bomem Inc., 585 Boul. Charest est, Suite 300, Québec, Qc, Canada G1K 9H4

## ABSTRACT

SpIOMM is an imaging Fourier transform spectrometer designed to obtain the visible range (350 – 850 nm) spectrum of every light source in a circular field of view of 12 arcminutes in diameter. It is attached to the 1.6-m telescope of the Observatoire du Mont Mégantic in southern Québec. We present here some results of three successful observing runs in 2007, which highlight SpIOMM's capabilities to map emission line objects over a very wide field of view and a broad spectral range. In particular, we discuss data cubes from the planetary nebula M27, the supernova remnants NGC 6992 and M1, the barred spiral galaxy NGC7479, as well as Stephan's quintet, an interacting group of galaxies.

**Keywords:** Fourier transform spectroscopy, hyperspectral imagery, planetary nebulae, supernova remnants, galaxies

## 1. INTRODUCTION

SpIOMM is an imaging Fourier transform spectrometer (FTS) attached to the 1.6-m telescope of the Mont Mégantic Observatory in southern Québec. Its early development phase was presented by Grandmont et al. (2003)[1] and Bernier et al. (2006)[2]. The SpIOMM instrument concept is based on the imaging version of the well known Fourier Transform Spectrometer (FTS). It is best described as an Integral Field Spectrometer (IFS) but it is fundamentally different from other known IFS since it does not rely on dispersive spectroscopy. It presents the particularity of being able to use all available detector pixels for imaging different field elements which explains its extremely wide field capabilities compared to other IFS. Its optical design is analog to the ones of classical cameras or focal reducers: a few lenses in line with an array detector (see Bernier et al., these proceedings[3], and Grandmont et al. 2008[4] for a recent, updated technical description of SpIOMM).

Spectral information is obtained by acquiring multiples panchromatic images of the scene that are modulated in intensity in a precisely controlled fashion. A Michelson interferometer inserted in series within the camera optical design modulates the image intensity from total dark to total brightness according to the optical path difference (OPD) between the two interferometer arms. This OPD being a function of wavelength, the resulting image intensity is the sum of all individual wavelength contributions. A recording of these panchromatic images generated at equally spaced OPD positions creates the raw data cube from which spectral information can be extracted. A Fourier transform calculation performed on a given pixel intensity recording across the multiple images of the raw datacube leads to a spectrum. Repeating this calculation for every pixel of the detector leads to a 3D spectral data set, producing one spectrum for every pixel of the detector (Figure 1). Spectral images are thus obtained without the need for image reconstruction, which increases the photometric and astrometric precision over standard IFS systems. Dual port interferometers provide access to the two complementary outputs. Their subtraction reconstructs the total incident flux of the interferogram and cancels the continuous term while their sum provides the total intensity of the sources, useful to monitor the sky transmittance over the data acquisition time.

---

[*] ldrissen@phy.ulaval.ca; phone 1 418 656 2131 x-5641; fax 1 418 656 2040

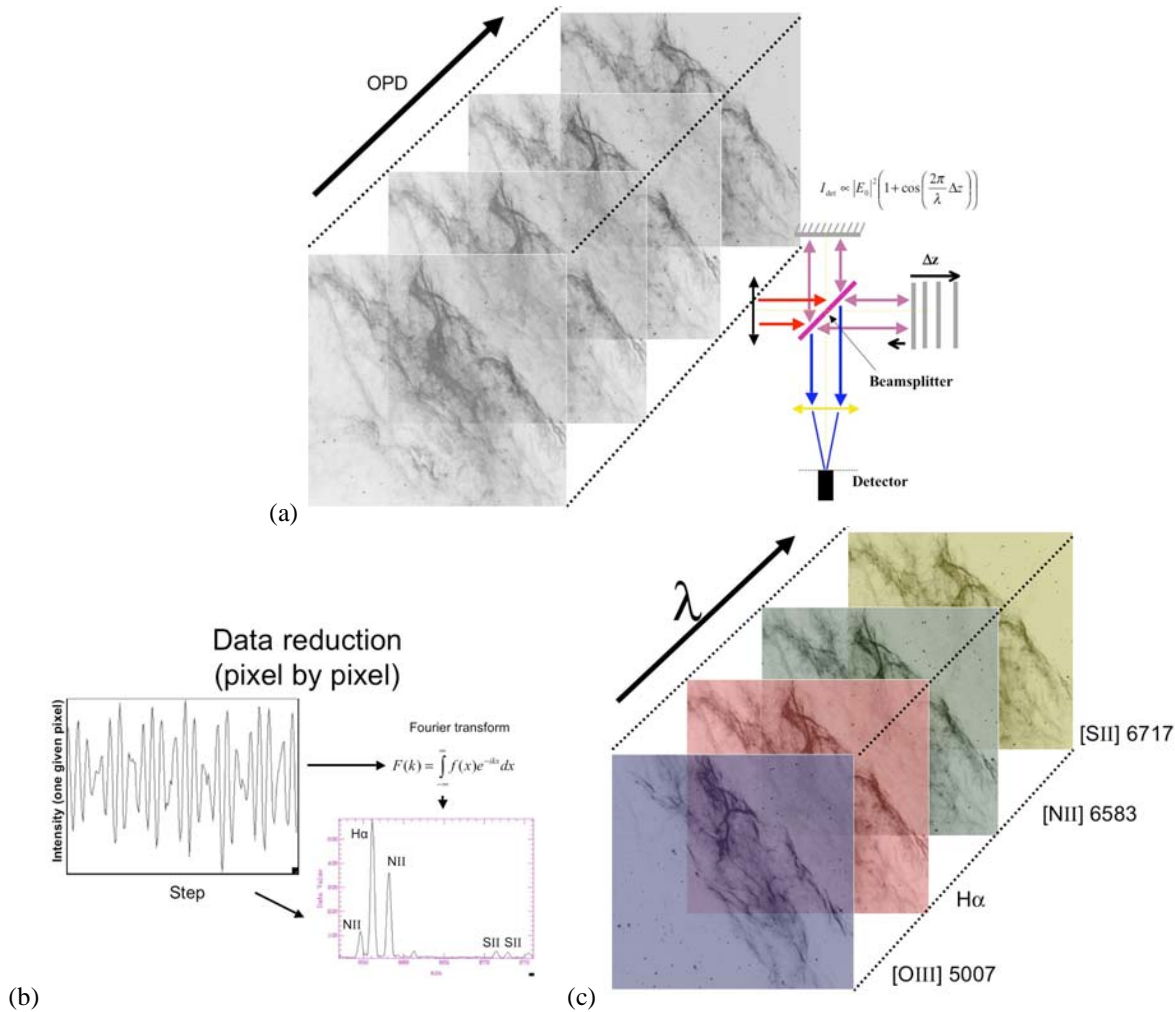

Fig. 1 Data acquisition with SpIOMM. (a) By scanning the Optical Path Difference (OPD) of the interferometer and taking images at every step, one gets a datacube. (b) For a given pixel, the recorded intensity varies as a function of the OPD with a pattern that depends on the spectral content of the source; for example, a monochromatic laser beam would produce a sinusoidal pattern. A Fourier transform of the signal produces a spectrum for every pixel in the image. (c) After Fourier transforming every interferogram (1.7 million), one gets a spectral datacube (RA, DEC, $\lambda$) from which monochromatic images corresponding to the emission lines of interest can be extracted. The data shown here come from a cube of the supernova remnant NGC 6992 (see Figure 4).

The light collection power of an Imaging FTS is also very high in dual output configuration given that the number of optical surfaces added by the interferometer can be as low as 1 mirror and 1 beamsplitter. The grating efficiency is analogously replaced by the modulation efficiency of the interferometer that would typically reach values higher than 80% over most of the instrument waveband. Hence, all the photons that are not absorbed by the camera lenses, the interferometer mirror or the beamsplitter are captured by either one of the detectors ; a by-product of the data cube is a very deep panchromatic image of the scene (right-hand panel of Figure 6). The total instrument efficiency can reach peak values of more than 65% including quantum efficiency. The readout time of the detector, typically around 5 seconds, reduces the overall efficiency of the instrument in the case of bright sources. The readout noise of the detector (around 5 electrons) also limits the depth of the individual exposures. Both limitations can be eliminated with the new generation of detectors such as those in use in some Fabry-Perot systems[5].

## 2. OBSERVATIONS WITH SPIOMM

Currently, only one output port is used on SpIOMM. The detector is a blue-sensitive, liquid nitrogen-cooled (to -120 $^{o}$C) Princeton Instruments VersArray CCD, with 1340 x 1300 pixels covering a field of view of 12 arcminutes x 12 arcminutes. Only the central, circular region with a radius of 6 arcminutes is non-vigneted, but the outer regions are still useful despite optical aberrations and a small decrease in sensisivity (see Figure 3 for a full field of view). The CCD is read at 100 kHz to reduce the readout noise and the pixels are binned 2x2 during readout to increase the S/N ratio and reduce the readout time.

As we will see in the next section, the « niche » we have chosen for SpIOMM is the mapping of emission lines in extended gaseous regions in the nearby universe : nebulae in the Milky way, and large gas-rich galaxies. While the observations could be done without any filter, thereby covering the entire visible range (from 350 nm to 850 nm), we take advantage of the fact that our prefered targets emit most of their flux in a series of emission lines clustered around 500 nm and 660 nm (rest wavelength). The observations are therefore performed in two steps, with a blue (450 – 520 nm) and a red (650 – 680 nm) filter to cover the most intense and diagnostic-rich emission lines (listed in Table 1). The use of filters also significantly increases the contrast between the targets (nebulae) and the underlying continuum sources (stars) and dramatically reduces the background sky intensity, especially when the Moon is up. Table 1 summarizes the filter settings. The most important criterium to define the spectral resolution is the capacity to unambiguously separate the lines from the [SII] doublet (whose ratio is a good indicator of the gas density). Such a resolution also ensures a clear separation of the [NII] doublet from the strong Hα line. On the red part, the line's full width at half maximum corresponds to ~ 130 km/s, but the centroid of each line can be determined with a precision of about 10 – 20 km/s, depending on the line strength, therefore allowing crude kinematical analysis of nebulae and rotation curves of galaxies. Exposure times vary from 7 seconds per step for the observations of bright nebulae in the red filter to 60 seconds per step for the observations of galaxies in the blue filter. A typical data cube therefore requires a total exposure time between one and five hours. During four observing runs in 2007 – 2008, we have obtained about 30 datacubes of galactic HII regions, planetary nebulae, Wolf-Rayet ring nebulae, supernova remnants, nearby spiral galaxies and groups of interacting galaxies.

Table 1. Filter settings

| Filter | Blue (450 – 520 nm) | Red (650 – 680 nm) |
|---|---|---|
| Emission lines ($\lambda_0$, nm) | He II 468.6 | [NII] 654.8 |
| | Hβ 486.1 | Hα 656.3 |
| | [OIII] 485.9 | [NII] 658.4 |
| | [OIII] 500.7 | He I 667.8 |
| | | [SII] 671.7 |
| | | [SII] 673.1 |
| Resolution ($\Delta\lambda$, nm) | 0.9 | 0.3 |
| Number of steps | 180 | 325 |

## 3. SCIENTIFIC RATIONALE AND SOME RESULTS

While the number of scientific programs for this type of instrument is potentially huge (from the study of individual stars in local star clusters to the search for high-redshift Ly-α emitters), we chose to illustrate its scientific potential by advocating the enormous benefits provided by a systematic, complete 3D mapping of extended emission-line sources such as local HII regions, nearby galaxies and distant galaxy clusters. We also present in this section some examples of data obtained at the Mont Mégantic Observatory, during the 2007 observing season.

### 3.1 Structure of Galactic HII regions

This project is argued along three different but complementary aspects of the physics of HII regions: the gas thermodynamics, the determination of relative gas abundances and the impact of shock fronts originating from the massive stars' stellar winds on the ionisation structure. Despite significant progress made in recent years on the first two subjects, there remains a very severe uncertainty: the small and large scale temperature inhomogeneities in HII regions are almost completely unknown. To our knowledge, the study of the third aspect is entirely new.

Photoionized nebulae are essential to determine the chemical composition of the ISM in galaxies, including the primordial helium abundance. For oxygen, the basic method is to assume an isothermal nebula and use a single value for the electronic temperature, derived directly from specific line ratios like [OIII] 495.9 + [OIII] 500.7)/[OIII] 436.3 or [NII] 654.8 + [NII] 658.4)/[NII]575.5. All semi-empirical models which rely on other line ratios are calibrated using abundances obtained from the ``direct'' method. It can be argued, however, argued that the direct method is unreliable since it is assumed that temperature inhomogeneities are small in nebulae. This problem, identified and quantified in the 1960's[7], has been fully studied theoretically via photoionization models[8,9,10]. The rms temperature fluctuations on the surface of the HII regions (noted $t^2$) can be estimated observationally[11]. The effect of these fluctuations on the abundances can then be evaluated through photoionization models. Stasinska's models[6] show that abundances can be underestimated by up to a factor of 2 if $t^2 \sim 0.04$. Although some observations have been performed, there is still a lot of controversy about actual temperature fluctuations in HII regions. For example, an upper limit of $t^2 \sim 0.015$ for galactic HII regions[12] and $t^2 = 0.001$ in M42[13]. In the case of M42 however, very different results have also been obtained with $t^2 \sim 0.055$[11]. Part of this discrepancy origins from the poor spatial sampling covered in these nebulae with long-slit spectrographs. Thus, despite its significance for abundance determination (galactic, extragalactic and cosmological) the problem of temperature fluctuations in HII regions is still not solved while its importance is well established.

Hence evidence exists for temperature variations in HII regions however its mapping is sorely missing while it would provide needed information to improve the dynamical modeling of these objects[14,15]. So far every derivation of the fluid dynamics equation either with or without shocks have been done using isothermal conditions. If this condition is not supported on large scales by observations a serious reassessment of the physical conditions would be needed. In fact, HII regions are not simple photoionized nebulae with an homogeneous distribution of gas and monotonous velocity gradient.

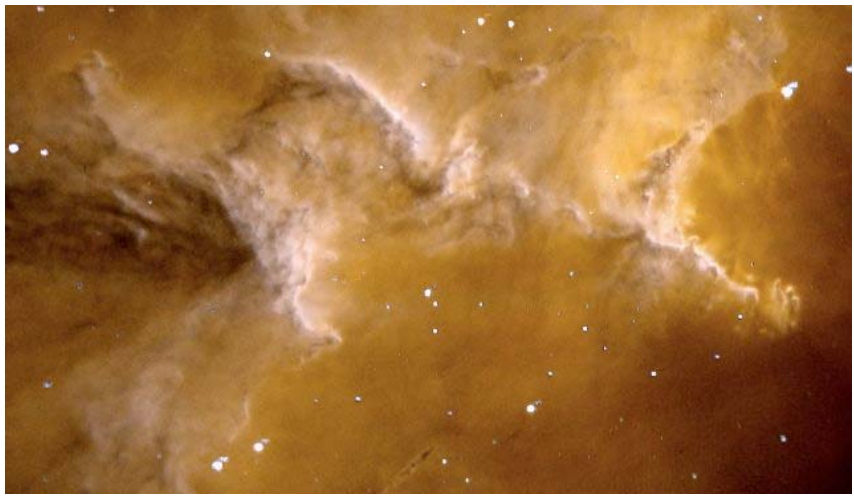

Fig. 2 Composite image (H$\alpha$ + [NII] 658.4) of a section of the HII region W4 (10' x 5'), located in the Persus arm of the Milky Way and ionized by massive OB stars in the IC 1805 cluster. This image, extracted from an SpIOMM data cube, shows the photodissociation interface region between the cold, dense molecular cloud, and the hot medium ionized by the massive stars.

Kinematical studies made by Joncas and others[16] proved the presence of turbulence in HII regions via the behavior of the radial velocity fluctuations. One of several statistical tools available to examine the characteristics of turbulence is the probability density function (PDF) increment. It checks for the presence of deviations from the famous Kolmogorov energy cascade where energy from the fluid flow is transferred from the large turbulent eddies to the smallest ones permitted by viscosity. These deviations are called intermittency. It describes the fact that turbulent activity can be confined at the end of the cascade to a small fraction of space leading to an uneven distribution in space and time of regions of large vorticity (shear) and dissipation. This energy dumping translates into local areas of increased temperature. Observationnally these areas are detected through a non-gaussian behavior of the PDF increments. Such behavior was detected in the HII regions Sh212 and Sh269. Electronic temperatures higher than the mean value of the nebula should be measured where intermittency is present. Is this the cause of the observed temperature fluctuations? Observers believe that the velocity fluctuations come from strong vortex tubes while theoreticians believe that they originate from the interface of 2-D shock fronts. Three-D spectroscopy and characterization of photoionisation will enable the distinction when helped with a proper knowledge of the velocity or density field. The latter being provided by the ratio of the [SII] line doublet.

Finally, 3-D spectroscopy of HII regions provides a unique way to analyze their ionisation structure. It has been known for a long time than massive stars emit copious winds. However their impact on the HII region structure lacks observational evidence. Shocks should be seen. Large, extended HII regions such as W4 (Figure 2) are prime targets for the kind of studies described above.

### 3.2 Nebulae around evolved stars

A meaningful link between local heavy element enrichment and the global chemical evolution of galaxies can only be established by detailed studies of individual wind-blown bubbles in our own galaxy. Winds of evolved stars, and their surrounding bubbles (planetary nebulae, Luminous Blue Variables and Wolf-Rayet ejecta, supernova remnants) are known to be globally enriched with products of nucleosynthesis. While planetary nebulae are ejected by low-mass stars, with slow (20-30 km/s) winds, LBVs and WRs are the late evolutionary stages of the most massive stars with wind velocities of hundreds to thousands of km/s.and A complete survey of abundance, density, temperature and kinematic measurements in nebulae surrounding individual evolved stars and ionizing clusters, looking for inhomogeneities in the distribution of processed material (primarily nitrogen and oxygen), which has never been undertaken because it requires wide-field spectroscopic mapping, will provide firm grounds for the interpretation of global galactic abundance studies. To illustrate this point, we have obtained blue and red datacubes of the planetary nebula M27 (Figure 3) and extracted maps of all relevant emission lines. We show in Figure 3 a ratio map of [NII] 658.4 / H$\alpha$ 656.3 which demonstrate the non-homogeneity of the physical conditions inside the nebula, as well as a BPT diagnostic diagram[17]. This type of diagram is frequently used to characterize the ionization source of distant, often unresolved, nebulae or galaxies and classify them (as HII regions, liners, AGNs, …). On most cases, every object in a BPT diagram is represented by a single point; each class of objects occupies a well-defined region. Truly wide-field integral spectroscopy such as provided by SpIOMM clearly shows that a single object actually samples a very wide range of physical conditions. Extended galactic supernova remnants will also be prime targets for SpIOMM. A classic example is the Cygnus loop, a 15 000 year-old supernova remnant spanning many degrees in diameter. A tiny fraction of this object has been mapped with the camera WFPC2 on the *Hubble Space Telescope*[18] to characterize the motion, structure and dynamical scale of the blast wave currently encountering the surrounding medium, in the northeastern part of the nebula. We have begun a complete mapping of the supernova remnant with SpIOMM is order to characterize this important object in its entirety. Figure 4 depicts some characteristics of a single red datacube, showing the unusually strong [SII] lines. This cube illustrates that SpIOMM can be used at the same time as an imager with a set of "perfect" narrow-band filters (note in Figure 4 how well separated all the emission lines are from their neighbors) as well as a low-resolution, very-wide field spectrograph: the right panel of this figure illustrates the global kinematics of the nebula showing well-defined filaments approaching and receding from us.

The case of another, much younger supernova remnant, M1 (also known as the Crab nebula), is particularly interesting as it illustrates the full power of SpIOMM: because of the presence of shocks, the forbidden lines of [NII] 654.8 and 658.4, as well as the [SII] 671.7, 3.1 doublet are almost as strong as the (usually strongest) H$\alpha$. Moreover, the gas is globally expanding at a velocity of up to ~ 1400 km/s. Therefore, up to 10 emission lines can be seen in regions where an approaching and a receding filament are superimposed on the line of sight (Figure 5) . Passing through the nebula mapped with SpIOMM with a visualizing software such as karma (http://www.atnf.csiro.au/computing/software/karma/) is a mesmerizing experience!

Ground-based and Airborne Instrumentation for Astronomy II, SPIE, Marseille, june 2008

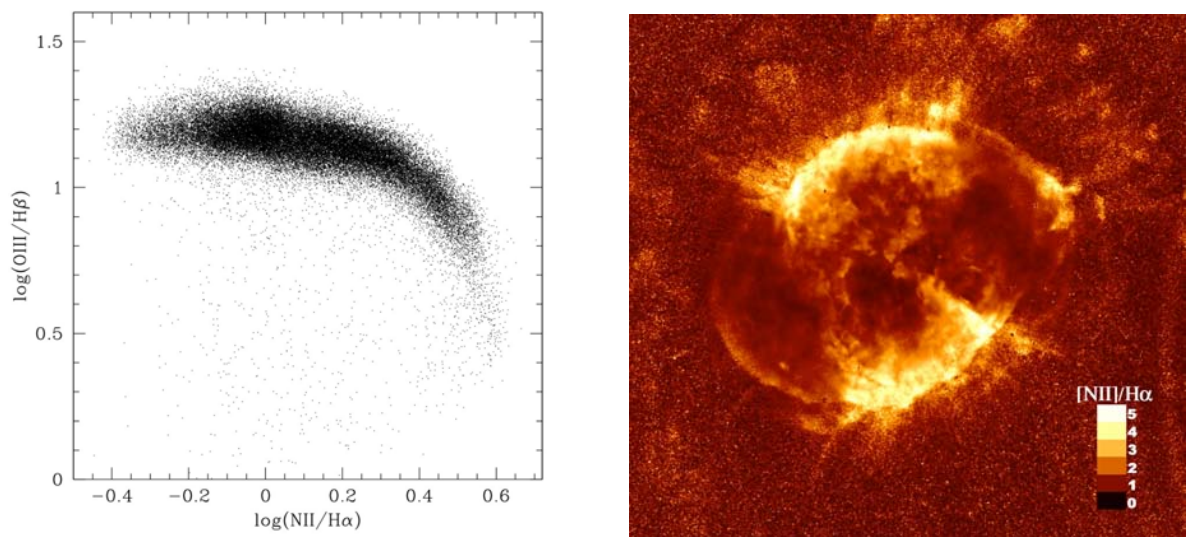

Fig. 3 (left) Diagnostic diagram of the planetary nebula M27; each point represents line ratios for a single 1.1 arcsecond pixel from the data cube. (right) Image of the [NII] 658.4 / Hα 656.3 line ratio from the same cube. Notice the large ratio, characteristic of shocks, at the outskirts of the inner bubble, as well as the periphery. The identification of the individual points on the diagnostic diagram with precise location on the image provides important constraints for the modelisation of the nebula.

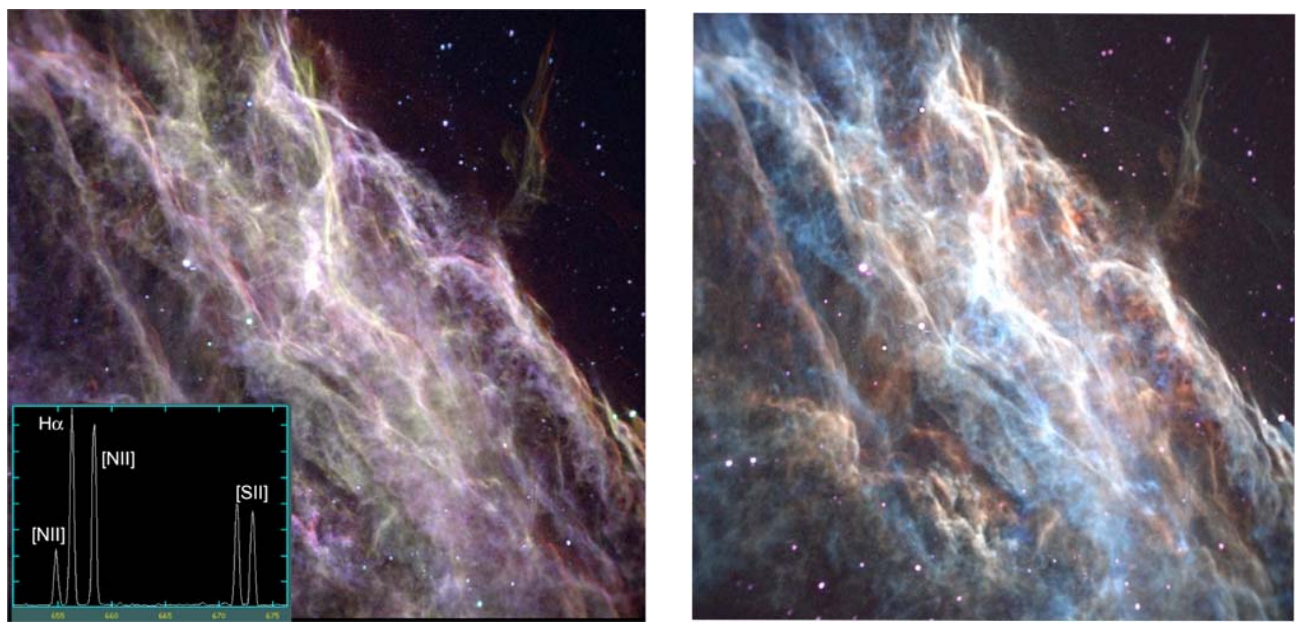

Fig. 4 (Left) Composite image (Hα + [NII] 6584) of a section of the supernova remnant NGC 6992 (part of the Cygnus Loop) extracted from a red (650 – 680 nm) data cube. The insert shows one of the 450 000 spectra from this cube. The most prominent emission lines are [NII] 654.8 nm, Hα 656.3 nm, [NII] 658.4 nm and the [SII] doublet at 671.7 and 673.1 nm. The very high [NII]/ Hα and [SII]/ Hα ratios are unusual in normal ionized nebulae; they are characteristic of shock regions and are common in supernova remnants. (Right) Velocity map in Hα, derived from the same data cube; the velocities range from -10 km/s (coded as blue) to +40 km/s (red).

Since the typical spectrum observed in the Crab nebula is largely subject to overlapping due to the Doppler shift of the different filaments on the line of sight, we have to proceed in a very systematic way to identify the elements that emit each line. The technique used here is similar to the one used by Čadež et al. 2004[19] in their study of the Crab nebula. Here we make an integral in which the spectrum is convolved with a Gaussian kernel adjusted to the resolution of the observed data for each line of the theoretical spectrum:

$$w_i(v) = \int S(\lambda) \exp\left(-\frac{(\lambda - \lambda_i(1 - v/c))^2}{2\sigma^2}\right)$$

where $S(\lambda)$ is the measured spectrum, $\lambda_i$ is the theoretical wavelength of each line, $v$ is the velocity (along the line of sight, with respect to the observer, of the emitting element, c is the speed of light and σ is the characteristic width of the gaussian kernel. So, $w_i(v)$ corresponds to the probability of finding the emitting element at a given velocity $v$ for each line i. In the case of the red (650 – 680 nm) datacube, i ranges from 1 to 5 and identify the lines listed in Table 1 with the exception of He I.

Since the [NII] and [SII] lines are each one doublet emission, the combined probability of finding the emitting element at a certain speed is much more reliable. So we define the total probability to find some matter at a given velocity $v$ to be the sum of the combined probability of each doublet except for the Hα, which is combined with the mean of the probability of the other elements:

$$\Omega(v) = w_1(v)w_2(v) + w_2(v)w_3(v) + w_4(v)w_5(v)$$

So, with this information, we can now locate the position of maximum probability to find the gas element and then adjust a gaussian to each observed line. To minimize the possible divergence of the fitting algorithm, only the amplitudes are considered as variable parameter since the velocities have already been found by the maxima of the $\Omega(v)$ function.

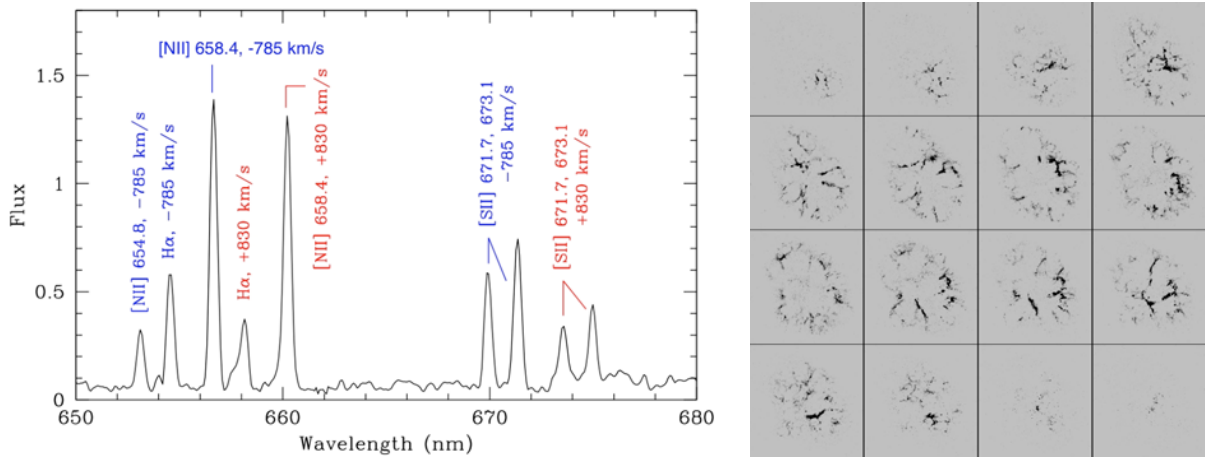

Fig. 5 (left) Typical spectrum (average of 10 pixels) of the Crab nebula, a young and rapidly expanding supernova remnant. Two groups of lines are clearly seen, a typical signature of two overlapping filaments on the line of sight. Note that the redshifted [NII] 654.8 nm line, not indicated on the graph, is superimposed on the blueshifted [NII] 658.4 nm component. (right) Montage of spectral channels from the Crab datacube (average of 10 channels per panel), ranging in velocities from -1400 km/s (approaching; top left panel) to + 1400 km/s (receding; bottom right panel), for the [NII] 658.4 nm emission line. Each panel is 7 arcminutes on a side.

### 3.3 Nearby galaxies

Powerful constraints on models of galactic chemical evolution, on the star formation histories of galaxies and on the dynamical processes that transform them can be derived from accurate and homogeneous determinations of (a) chemical abundances in individual gaseous nebulae; (b) the distribution of their stellar populations in terms of age and metallicity, and (c) the gaseous and stellar kinematics. So far, systematic studies between chemical properties (central abundance or radial abundance gradient), and other parameters have been conducted for a small sample (~ 50) of spirals. The effects of the morphological type, the presence of bars, and environment have all been studied to some extent. However, although informative, these studies suffer from several severe limitations: (1) Abundances are generally derived from a few individual HII regions in each galaxy ; (2) Long-slit, MOS and IFU spectroscopy is limited to a tiny fraction of the total surface of their targets; (3) There is a lack of uniformity in the calibration methods used to determine the abundances; (4) A limited sample of galaxies across the Hubble sequence has been well observed; (6) There are still large uncertainties arising from diverse spectrophotometric techniques (narrow slits, narrow-band imaging). Moreover, most studies so far, performed with slit spectrographs, have concentrated on global properties of stellar ejecta or abundance gradients in galaxies, thereby neglecting possible small-scale variations caused by multi-phase stellar wind (individual stars) or localized enrichment by starburst clusters in peculiar evolutionary stages. SpIOMM, or its potential successor SITELLE (see section 4 below) will be an ideal instrument to conduct a systematic study of abundances in a large sample of nearby galaxies and thus easily detect evidence for small-scale enrichments and establish conditions under which they take place. The possibility to study the multiple emission line ratios and kinematics for hundreds of HII regions simultaneously in each individual galaxy would completely remove some of the limitations mentioned above. With the availability of several lines *simultaneously* (e.g. [OII] 372.7, 732.0, 733.0 nm ; [OIII] 436.3, 495.9, 500.7 nm; [NII] 575.5, 654.8, 658.4 nm; Balmer series) as opposed to Fabry-Perot or interference filter data which cover a single line, diverse abundance calibrations could then be applied and evaluated. Absorption lines due to the presence of underlying stars will be obtained simultaneously, thereby providing crucial information about older stellar populations, such as age and metallicity. With its high efficiency and large field of view, SpIOMM will be the ideal instrument to carry a large program on the chemical abundances, kinematics and stellar populations of a significant sample of nearby galaxies across the Hubble sequence. Since most of the conclusions drawn so far on the chemical composition of galaxies are based on restricted and inhomogeneous data, such a dataset would revolutionize the study of galaxy chemical evolution.

We have obtained datacubes of half a dozen galaxies so far, including the barred spiral NGC 7479 and the almost edge-on spiral NGC 7331 (Figure 6). Ratios of the brightest emission lines in the red filter can be obtained for a large number of HII regions across the disk, enven in the presence of a large velocity field. Emission lines in edge-on galaxies are very difficult to map with traditional techniques such as narrow-band filters because the rotational velocities push the lines in and out of the filters which prevent accurate flux measurements.

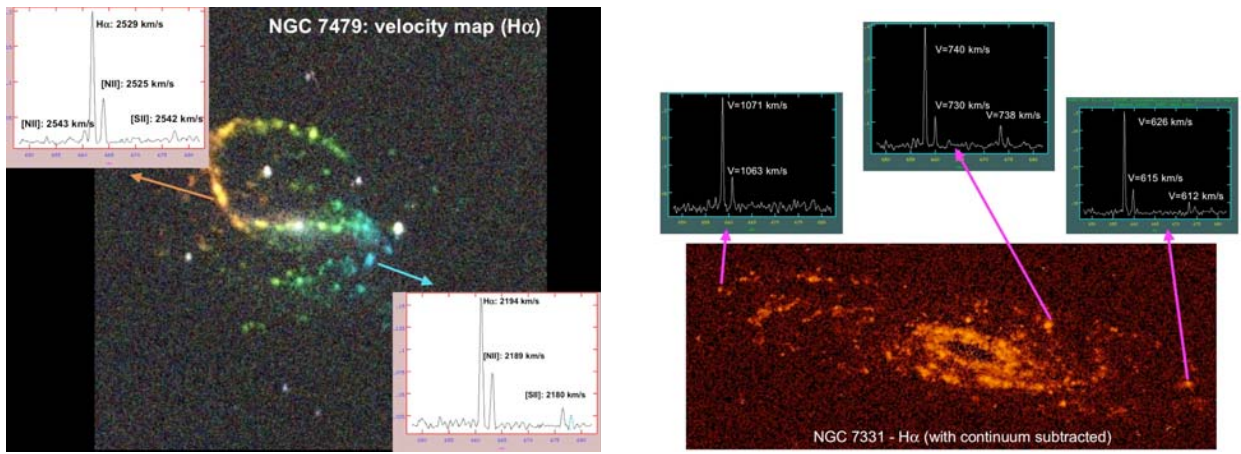

Fig 6. (Left) Velocity map of the barred spiral galaxy NGC 7479, spanning more than 350 km/s. (Right) Pure Hα map of the almost edge-on spiral NGC 7331, spanning the entire velocity range inaccessible with a traditional Fabry-Perot system.

### 3.4 Galaxies across the Universe

Because SpIOMM can be used without predefined filter in the whole visible waveband, it will be extremely useful to study the intermediate-to-high-z Universe in a relatively unbiased way. It will, for example, allow detection of Ly-$\alpha$ emission in objects from $z = 2$ to 7. The power of SpIOMM, but most likely its successor attached to a larger telescope, in this type of study is that it will sample the redshift space uniformly on a wide field, allowing spectroscopic redshift determination and line profile analysis on several galaxies in a single observation. In the same vein, chemical abudance studies at intermediate redshift, or the study of the kinematics in the intergalactic gas in clusters, could also be performed. As an example, we show in Figure 7 images from a datacube of Stephan's quintette, a group of interacting galaxies at redshift $z \sim 0.02$ in front of which lie a nearby interloper, NGC 7320, a member of the same group as NGC 7331 (illustrated in Figure 6). The datacube clarly separates the star-forming HII regions in NGC 7320 from the more distant galaxies. The active galactic nucleus in the Seyfert galaxy NGC 7319 is also well detected.

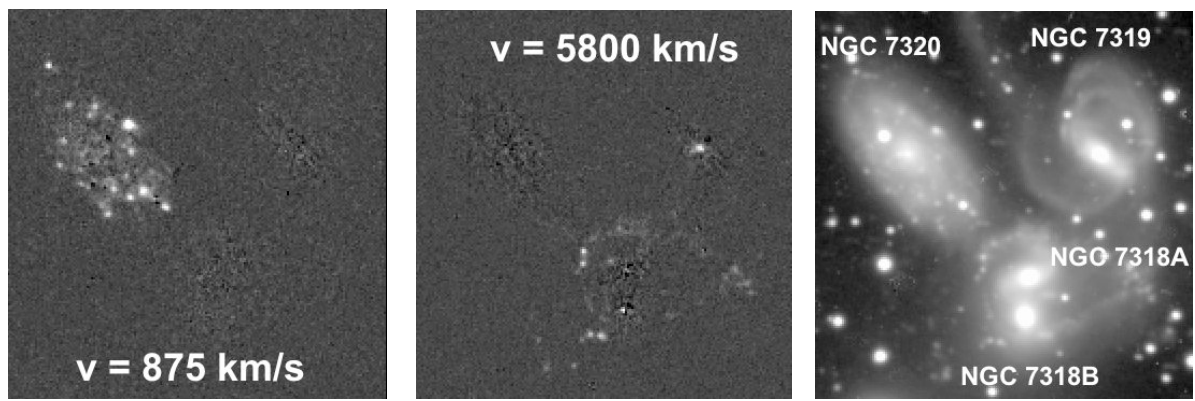

Fig 7. (Left and center) Slices extracted from the data cube of Stephan's Quintet, showing the ionized gas in the foreground galaxy NGC 7320 (average recession velocity of 875 km/s) and the HII regions from the highly disturbed spiral arms in NGC 7318A (average velocity of 5800 km/s with a high dispersion). Notice also, in the middle panel, the strong emission coming from the Seyfert nucleus of NGC 7319. The right-hand panel shows the deep panchromatic (650 – 680 nm) image resulting from the addition of all the interferograms.

## 4. SITELLE

We are presently conducting a feasibility study for a similar instrument to be attached to the Canada-France-Hawaii 3.6-m telescope, dubbed SITELLE (Spectromètre Imageur à Transformée de Fourier pour l'Etude en Long et en Large de raies d'Emission). Having learned from the design and use of SpIOMM, some significant improvements will be included in the new concept, including the servo mechanism and the detector technology. Taking into account the larger surface of the primary mirror, and the improved modulation efficiency and detector quantum efficiency, we estimate that SITELLE will be about 20 times more efficient than SpIOMM and will allow the mapping of fainter object as well as fainter emission lines.

## ACKNOWLEDGEMENTS

We would like to thank Jean-Pierre Maillard, who has pioneered the use of FTS at the Canada-France-Hawaii telescope, for his continuous support since the early stages of the development of SpIOMM, as well as ABB Bomem, for its technical support. LD acknowldeges funding by the Canadian Foundation for Innovation, the Canadian Space Agency, Québec's FQRNT, Canada's NSERC, and Université Laval.

# REFERENCES

[1] Grandmont, F., Drissen, L., and Joncas, G., "Development of an Imaging Fourier Transform Spectrometer for Astronomy", in Specialized Optical Developments in Astronomy. Edited by Atad-Ettedgui, E., D'Odorico, S., Proceedings of the SPIE, Volume 4842, pp. 392-401 (2003).

[2] Bernier, A.-P., Grandmont, F., Rochon, J.-F., Charlebois, M., and Drissen, L., "First results and current development of SpIOMM: an imaging Fourier transform spectrometer for astronomy"

[3] Bernier, A.-P., Charlebois, M., Drissen, L., & Grandmont, F., "Technical improvements and performances of SpIOMM: an imaging Fourier transform spectrometer for astronomy", these proceedings.

[4] Grandmont, F., Drissen, L., Bernier, A.-P., Charlebois, M. "SpIOMM: An imaging Fourier transform spectrometer for astronomy", The Astronomical Journal, submitted (2008).

[5] Dicaire, I., Carignan, C., Amram, P., Marcelin, M., Hlavacek-Larrondo, J., de Denus-Baillargeon, M.-M., Daigle, O., Hernandez, O., "Hα kinematics of the Spitzer Infrared Nearby Galaxies Survey - II", AJ, 135, 2038 (2008).

[6] Stasinska, G., "How reliable are abundances from nebular spectra?", in Abundance profiles: diagnostic tools for galaxy history, ASP Conf. Ser. 147, p. 142 (1998)

[7] Peimbert, M., "Temperature Determinations of H II Regions", ApJ, 150, 825 (1967).

[8] Garnett, D., "Electron temperature variations and the measurement of nebular abundances", AJ, 103, 1330 (1992).

[9] Steigman, G., Viegas, S. M., Gruenwald, R., "Temperature Fluctuations and Abundances in H II Galaxies", ApJ, 490, 187 (1997).

[10] Mathis, J. S., Torres-Peimbert, S., Peimbert, M., "Temperature and Density Fluctuations in Planetary Nebulae", ApJ, 495, 328 (1998).

[11] Walter, D. K., Dufour, R. J., Hester, J. J., "CNO abundances and temperature fluctuations in the Orion Nebula", ApJ, 397, 196 (1992).

[12] Shaver, P. A., et al., "The galactic abundance gradient", MNRAS, 204, 53 (1983).

[13] Liu, X.-W., Barlow, M. J., Danziger, I. J,, Storey, P. J., "Balmer Discontinuity Temperatures in the Orion Nebula", ApJ, 450, L59 (1995).

[14] Binette, L., Ferruit, P., Steffen, W., Raga, A. C., "Relation between Source and Temperature Fluctuations in Photoionized Nebulae", RMxAA, 39, 55 (2003).

[15] Copetti, M. V. F., "Electron temperature fluctuations in H II regions. The feasibility of t2 estimates from point-to-point observations", A&A, 453, 943 (2006).

[16] Joncas, G., "Turbulence in HII regions: New results", in Interstellar Turbulence, Proceedings of the 2nd Guillermo Haro Conference. Edited by Jose Franco and Alberto Carraminana. Cambridge University Press, p. 154 (1999).

[17] Baldwin, J. A., Philips, M. M., Terlevich, R., "Classification parameters for the emission-line spectra of extragalactic objects", PASP, 93, 5 (1981).

[18] Blair, W. P, Sankrit, R., Raymond, J. C., "Hubble Space Telescope Imaging of the Primary Shock Front in the Cygnus Loop Supernova Remnant", AJ, 129, 2268 (2005).

[19] Čadež, A., Carramiñana, A., Vidrih, S., "Spectroscopy and Three-Dimensional Imaging of the Crab Nebula", ApJ, 609, 797 (2004).